\documentclass[twocolumn,showpacs,aps,prb,floatfix]{revtex4}

\usepackage[dvips]{graphicx}
\usepackage{dcolumn}
\usepackage{bm}

\begin{document}

\preprint{APS/123-QED}

\title{
Magnetoresistance and spin polarization in the insulating regime of a Si two-dimensional electron system
}

\author{Mitsuaki Ooya}
\author{Kiyohiko Toyama}
\author{Tohru Okamoto}

\affiliation{
Department of Physics, University of Tokyo, 7-3-1 Hongo, Bunkyo-ku, Tokyo 113-0033, Japan 
}%

\date{\today}

\begin{abstract}
We have studied the magnetoresistance in a high-mobility Si inversion layer down to low electron concentrations at which the longitudinal resistivity $\rho_{xx}$ has an activated temperature dependence. The angle of the magnetic field was controlled so as to study the orbital effect proportional to the perpendicular component $B_\perp$ for various total strengths $B_{\rm tot}$. A dip in $\rho_{xx}$, which corresponds to the Landau level filling factor of $\nu=4$, survives even for high resistivity of $\rho_{xx} \sim 10^8~\Omega$ at $T= 150~{\rm mK}$. The linear $B_{\rm tot}$-dependence of the value of $B_\perp$ at the dip for low $B_{\rm tot}$ indicates that a ferromagnetic instability does not occur even in the far insulating regime.

\end{abstract}

\maketitle

\section{\label{sec:level1}introduction}
There has been great attention to the fundamental properties of strongly correlated two-dimensional (2D) electron or hole systems in the last decade, due in part to the discovery of the zero-magnetic-field metal-insulator transition (MIT). \cite{Kravchenko1994,Abrahams2001,Kravchenko2004}
The strength of the Coulomb interaction between the carriers is characterized by the Wigner-Seitz radius $r_s$, which is equal to the ratio of the Coulomb energy per electron to the Fermi energy.
In the Fermi liquid theory, the effective mass $m^ *  $, the effective $g$-factor $g^ *  $ and the spin susceptibility $\chi ^ *   \propto g^ *  m^ *  $ are renormalized by $r_s $, and they are expected to be enhanced as $r_s $ increases. \cite{Iwamoto1991,Kwon1994,Chen1999}
The dimensionless parameter $r_s$ can be written as $r_s = \pi^{1/2} (e/h)^2 (m_b/\kappa \varepsilon_0) N_s {}^{-1/2}$ and controlled by changing the electron concentration $N_s$, where $m_b$ is the band mass and $\kappa$ is the average dielectric constant.
The enhancement of the spin susceptibility with increasing $r_s$ (decreasing $N_s$) has been observed in various 2D electron systems (2DESs) formed in Si metal-oxide-semiconductor field-effect-transistors (MOSFETs), \cite{Fang1968,Okamoto1999,Pudalov2002} GaAs/AlGaAs, \cite{Zhu2003} Si/SiGe \cite{Okamoto2004} and AlAs/AlGaAs \cite{Shkolnikov2004} heterostructures.
For 2DESs without disorder, the ferromagnetic transition is expected to occur at $r_s \sim 26$ before the Wigner crystallization at $r_s \sim 35$. \cite{Attaccalite2002}
Recently, the divergence of $\chi ^ *$ at or near the MIT with $r_s \sim 9$ has been reported for Si-MOSFETs, \cite{Shashkin2001,Vitkalov2001}
while the results in Refs.~\onlinecite{Pudalov2002} and \onlinecite{Pudalov2001} do not support the occurrence of a ferromagnetic instability at the MIT.
Another 2D Fermi liquid system, $^{\rm{3}} {\rm{He}}$ absorbed on graphite, also shows a tendency for $\chi ^ *  $ to diverge as the system goes into localization. \cite{Casey2003,Tsuji2004}

A possible ground state in the insulating regime of high-mobility Si-MOSFETs is a pinned Wigner crystal (WC) or glass.
Pudalov {\it et al.} observed nonlinear dc conduction with a sharp threshold electric field in the insulating regime of Si-MOSFETs and attributed it to that of a pinned WC, \cite{Pudalov1993} while it was also discussed in terms of the single particle localization picture taking into account a Coulomb gap. \cite{Shashkin1994,Mason1996}
Chui and Tanatar found from their Monte Carlo studies that the WC can be stabilized at $r_s$ as low as 7.5 in the presence of a very small amount of disorder in the oxide layer. \cite{Chui1995,Chui1997}
In the WC at rather low $r_s$, exchanges among neighboring electrons are expected to occur frequently.
The amplitudes of several types of ring exchanges in Si-MOSFETs were calculated \cite{Okamoto1998,Katano2000} to be in the order of 0.1~K at $r_s \sim 8$ using the WKB method developed by Roger. \cite{Roger1984}
The strength of the ferromagnetic interaction, which arises from exchanges of an odd number of particles, is comparable to that of the antiferromagnetic interaction from exchanges of an even number of particles. \cite{Katano2000}
Furthermore, the valley degree of freedom in Si inversion layers makes the system more complicated. \cite{afterOkamoto1998}
It seems hard to predict theoretically the magnetic ground state of the WC in Si 2DESs.

In this work, we have performed systematic measurements of magnetoresistance of a high-mobility Si-MOSFET in order to study the electronic and spin states in the insulating regime.
The perpendicular component $B_{\perp}$ of the magnetic field was controlled by rotating the sample for various total strengths $B_{\rm tot}$ so as to investigate the orbital effect independently of the Zeeman effect.
We observed that a dip in the longitudinal resistivity $\rho_{xx}$, which corresponds to the Landau level filling factor of $\nu=4$, remains even in the far insulating regime where the Landau levels are expected to be smeared out.
The value of $B_\perp$ at the dip shows a linear $B_{\rm tot}$-dependence in the low $B_{\rm tot}$ region below a kink indicating the onset of the full spin polarization.
The results strongly suggest that a ferromagnetic instability does not occur in the insulating regime.

\section{sample and experimental method} 
We used a (001)-oriented Si-MOSFET sample with a peak electron mobility $\mu _{{\rm{peak}}}$ = 2.4 ${\rm{m}}^{\rm{2}} /{\rm{V s}}$ at $N_s= 4 \times 10^{15}~{\rm{m}}^{-2} $ and $T$ = 0.3 K. 
It has a Hall-bar geometry of total length 3 mm and width 0.3 mm. 
The estimated SiO${}_2$ layer thickness is $98~{\rm nm}$.
Standard DC four-probe techniques were used to measure $\rho_{xx}$ and the Hall resistivity $\rho_{xy}$. 
The potential probes were separated by 1.5 mm and the excitation voltage had been kept 0.4 mV to ensure that measurements were taken in the \textit{I-V} linear regime. 
The electron concentration $N_s $ was controlled by varying the gate voltage and determined from the Hall coefficient measured at $T$ = 3 K. 
The MIT is observed at the critical electron concentration $N_c  = 0.97 \times 10^{15}~{\rm{m}}^{-2}$ in the absence of the magnetic field.
It is estimated that $r_s = 8.4$ at $N_s  = N_c$ with $m_b = 0.19 m_0$ and $\kappa=7.7$. \cite{Ando1982}
The sample was mounted on a rotatory thermal stage in a dilution refrigerator together with a GaAs Hall generator and a ${\rm{RuO}}_{\rm{2}} $ resistance thermometer calibrated in magnetic fields. 
The rotatory thermal stage was cooled via a silver foil linked to the mixing chamber and the temperature was accurately controlled by a heater on the stage.

\section{experimental results and discussion}
Figure 1 shows the $B_ \perp $-dependence of the longitudinal resistivity $\rho _{xx} $ at $B_{\rm tot} = 9~{\rm T}$ for various $N_s$ with a wide range of $\rho _{xx} $.
\begin{figure}
\includegraphics{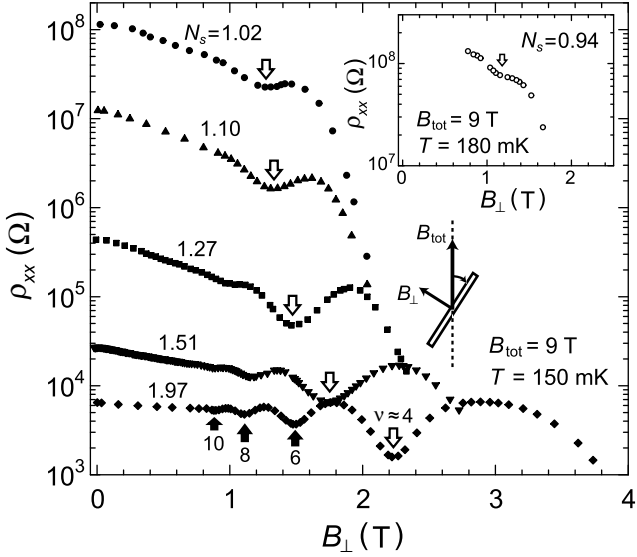}
\caption{
Longitudinal resistivity as a function of $B_\perp$ at $B_{{\rm{tot}}} = 9~{\rm T}$ and $T = 150~{\rm mK}$ for various $N_s $ indicated in units of $10^{15}~{\rm m}^{-2}$. 
The inset shows the data at $T = 180~{\rm mK}$ for $N_s =0.94 \times 10^{15}~{\rm m}^{-2}$.
}
\end{figure}
As discussed later, the electron spins are expected to be fully polarized at $B_{{\rm{tot}}} = 9~{\rm T}$ for these values of $N_s$. \cite{Okamoto1999}
For $N_s = 1.97 \times 10^{15}~{\rm m}^{-2}$, minima in $\rho_{xx}$ are observed at the Landau level filling factors $\nu \approx 4$, 6, 8 and 10, which correspond to the Shubnikov-de Haas (SdH) oscillations of the spin polarized system. 
Note that twofold valley degeneracy remains in Si(001) inversion layers.
As $N_s $ decreases, $\rho_{xx}$ drastically increases and the dips at $\nu \approx 6$, 8 and 10 are smeared out.
It is reasonable that the SdH oscillations disappear when a dimensionless parameter $\omega_c \tau$ becomes less than unity. Here $\omega_c$ is the cyclotron frequency and $\tau$ is the cyclotron scattering time.
This condition can be rewritten as $\rho_\parallel \geq h/ \nu e^2$ ($\approx \rho_{xy}$) if $\tau$ is replaced by the classical scattering time $\tau_c$ obtained from the resistivity $\rho_\parallel \equiv \rho_{xx}(B_\perp =0)$.
Measurements on low-mobility Si-MOSFETs \cite{Fang1977} and GaAs/AlGaAs 2DESs \cite{Harrang1985} have shown that $\tau$ is comparable or smaller than $\tau_c$.
Thus, in a simple picture, the SdH oscillations are not expected to appear for $\rho_\parallel \gtrsim 10^4~\Omega$.
However, the dip at $\nu \approx 4$ remains for very high resistivity up to $\rho_{xx} \sim 10^8~\Omega$.
Distinct $\rho_{xx}$ minima at $\nu \approx {\rm integer}$ in the insulating regime of high-mobility Si-MOSFETs have also been observed for $\nu \approx 1$ and 2 in perpendicular magnetic fields ($B_\perp = B_{\rm tot}$). \cite{Kravchenko1991,DIorio1992}
It is well known that the usual SdH oscillations originate from the $B_\perp$-dependence of the density of states at the Fermi level $\varepsilon_{F}$.
The longitudinal conductivity $\sigma_{xx}=\rho_{xx}/(\rho_{xx} {}^2+\rho_{xy} {}^2)$ has a minimum when $\varepsilon_{F}$ lies in a gap between Landau levels.
For $\rho_{xx} \ll \rho_{xy}$, a minimum in $\sigma_{xx}$ leads to a minimum in $\rho_{xx}$.
In the insulating regime with $\rho_{xx} \gg \rho_{xy}$, on the other hand, it leads to a maximum in $\rho_{xx}$, 
which is contrary to the experimental results for $\nu \approx {\rm integer}$.

Figure~2(a) shows the $B_\perp$-dependence of $\rho_{xx}$ in the insulating regime for various temperatures.
\begin{figure}
\includegraphics{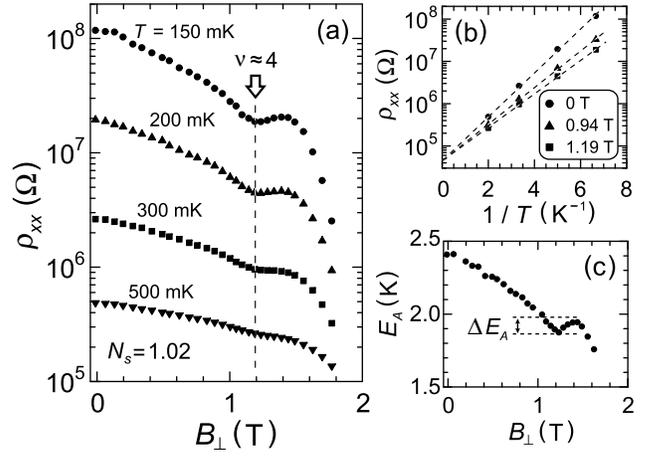}
\caption{
Data in the insulating regime for $B_{{\rm{tot}}}$ = 6 T and $N_s  = 1.02 \times 10^{15}~{\rm{m}}^{-2} $.
The electron spins are expected to be fully polarized. \cite{Okamoto1999}
(a) $B_\perp$-dependence of $\rho_{xx}$ for various temperatures. 
(b) Arrhenius plots of $\rho_{xx}$ for $B_ \perp   = $ 0 T, 0.94 T and 1.19 T (at the dip). 
(c) Activation energy as a function of $B_ \perp$.
}
\end{figure}
$\rho_{xx}$ decreases drastically with increasing temperature even for the minimum at $\nu \approx 4$ while $\rho_{xx}$ increases with $T$ for minima in the usual SdH oscillations. \cite{Pudalov2002}
While the dip at $\nu \approx 4$ in Fig.~2(a) is gradually smeared out as $T$ increases, the position of the dip does not depend on $T$.
Figure 2(b) shows Arrhenius plots of $\rho _{xx} $ for different $B_ \perp $. 
The dashed lines are least-square fits to the experimental data and represent $\rho_{xx} = \rho_0 \exp (E_{A}/2T)$.\cite{VRH}
The $B_\perp$-dependence of $\rho_{xx}$ at low temperature is attributed to a change in the activation energy $E_{A}$ rather than that in the prefactor $\rho_0$.
The obtained $E_{A}$ is shown in Fig. 2(c) as a function of $B_ \perp $. 
In the low $B_ \perp$ region, $E_{A}$ decreases almost linearly with increasing $B_ \perp$.
This might be associated with the delocalization effect of the magnetic field in the strongly localized regime.
$E_{A}$ shows a dip at $\nu \approx 4$.
Although it might be a trace of the Landau level formation, the depth of the dip of $\Delta E_{A} \sim 0.1~{\rm K}$ is much smaller than the Landau level spacing of $\hbar \omega_c = 8.5~{\rm K}$ at $B_\perp =1.2~{\rm T}$ expected from the band mass of $m_b =0.19m_0$.
At this stage, the origin of $\rho_{xx}$ or $E_{A}$ minima at $\nu \approx {\rm integer}$ observed in the insulating regime is not understood.
We have investigated $B_\perp$-dependence of $\rho_{xx}$ also for other 2D systems with high resistivity of $\rho_{xx} \gtrsim 10^6~\Omega$ at low temperatures down to $100~{\rm mK}$. \cite{Toyama}
Dips at $\nu \approx 1$ and 2 are observed for GaAs hole systems with $r_s \approx 10$, while an insulating GaAs electron system with small $r_s$ ($\approx 3$) only shows a broad minimum resulting from negative magnetoresistance at low $B_\perp$ and positive one at high $B_\perp$ owing to the shrinkage of the electron wave function. 
Such $B_\perp$-dependence of $\rho_{xx}$ in a GaAs 2DES was also observed in Fig.~2 of Ref.~\onlinecite{Jiang}. The collective motion of electrons might cause the dips in $\rho_{xx}$ at $\nu \approx {\rm integer}$ observed in the insulating regime of the strongly correlated 2D systems.

The values of $B_\perp$ at $\rho_{xx}$ minima depend on $B_{\rm tot}$.
Typical data in the low resistivity region are shown in Fig.~3(a).
The positions of the $\rho_{xx}$ minima at $\nu \approx 4$ and $6$ in the SdH oscillations shift toward low-$B_\perp$ side as $B_{\rm tot}$ decreases.
This can be explained as the result of a decrease in the fraction of ``spin-up'' electrons. \cite{Okamoto1999,Okamoto2004,Okamoto2000,Tutuc2001,Tutuc2002}
\begin{figure}
\includegraphics{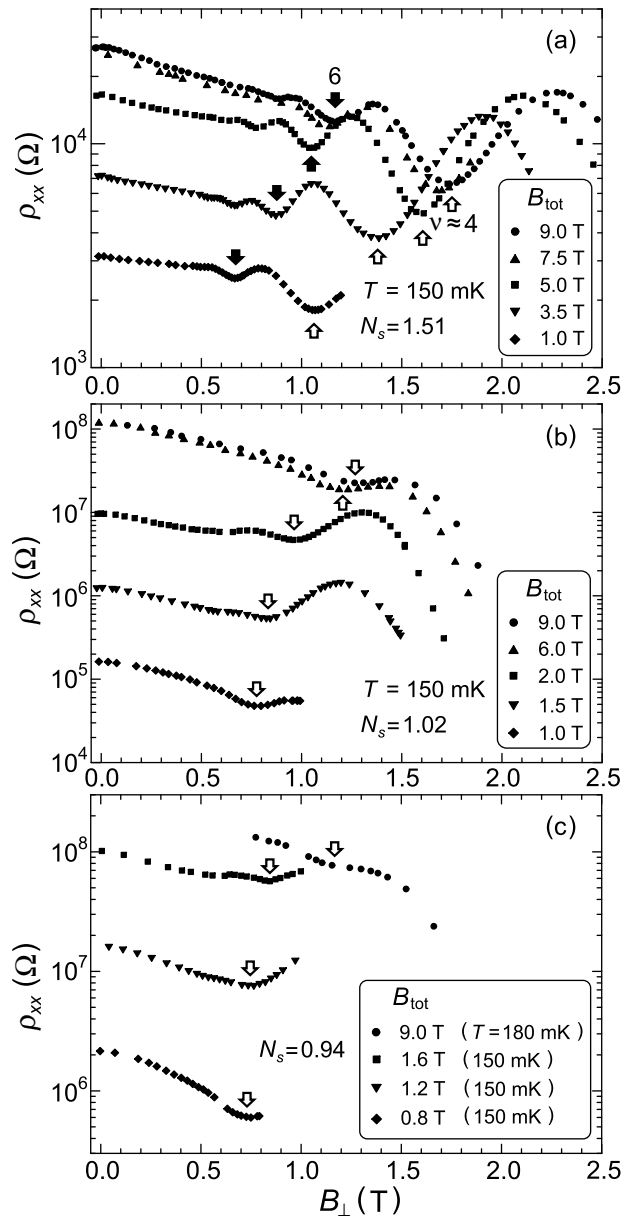}
\caption{
$B_\perp$-dependence of $\rho_{xx}$ for various $B_{\rm tot}$.
(a) Data in the low resistivity region with $N_s = 1.51 \times 10^{15}~{\rm m}^{-2}$.
(b)(c) Data in the higher resistivity region with $N_s = 1.02$ and $0.94 \times 10^{15}~{\rm m}^{-2}$, respectively.
}
\end{figure}
Similar behavior is observed in the insulating regime.
In Fig.~3(b) and (c), the data for $N_s = 1.02$ and $0.94 \times 10^{15}~{\rm{m}}^{-2}$ are shown, respectively.
The dip at $\nu \approx 4$ shifts toward low-$B_\perp$ side as $B_{\rm tot}$ decreases.
In Fig.~4, the positions of the $\rho _{xx} $ minima determined taking into account the negatively  $B_\perp$-dependent baseline are shown as $1/\nu_{\rm min} = e B_\perp / h N_s$ for different $N_s$.
Overall behavior for low $N_s$ is similar to that for high $N_s$, i.e., $1/\nu_{\rm min}$ increases almost linearly with $B_{\rm tot}$ before $B_{\rm tot}$ exceeds a critical value $B_c$ indicated by arrows.
We consider that $B_c$ corresponds to the onset of the full spin polarization of 2D electrons.
The dotted lines represent $\nu_\uparrow =4$ or 6 assuming that the spin polarization $P = 2 N_\uparrow /N_s -1$ increases linearly with $B_{\rm tot}$ for $B_{\rm tot} \leq B_c$.
Here $\nu_\uparrow$ and $N_\uparrow$ are the Landau level filling factor and the concentration of spin-up electrons, respectively ($\nu_\uparrow = h N_\uparrow / e B_\perp $).
The SdH oscillations depending on $\nu_\uparrow$ are also observed in a Si/SiGe sample with much lower resistivity. \cite{Okamoto2004}
Although the reason for the survival of the dip in the far insulating regime is unclear, the dotted line for $\nu_\uparrow = 4$ well reproduces the experimental results for $B_{\rm tot} \leq B_c$.
The $B_{\rm tot}$-dependent behavior for $B_{\rm tot} \leq B_c$ strongly suggests that the spin polarization is not completed and a ferromagnetic instability does not occur.
\begin{figure}
\includegraphics{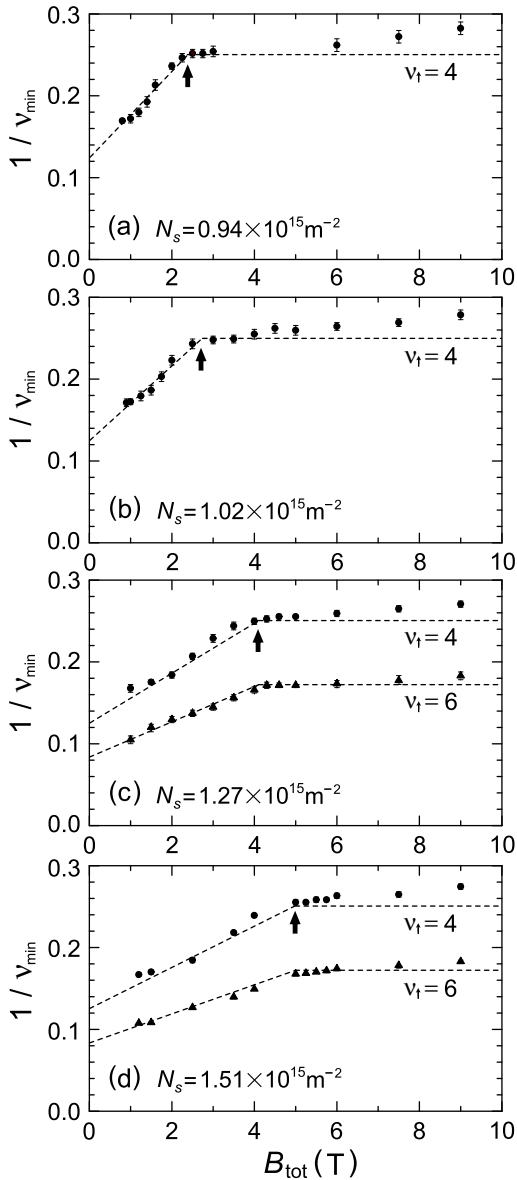}
\caption{
Position of $\rho _{xx} $ minima $1/\nu_{\rm min} = e B_\perp / h N_s$. Data for (a) $N_s  = 0.94 \times 10^{15}~{\rm{m}}^{-2}$, (b) $1.02 \times 10^{15}~{\rm{m}}^{-2} $, (c) $1.27 \times 10^{15}~{\rm{m}}^{-2} $ and (d) $1.51 \times 10^{15}~{\rm{m}}^{-2}$ are shown.
The arrows indicate $B_c$.
The dotted lines represent $\nu_\uparrow =4$ or 6 (see text).
}
\end{figure}

A gradual increase in $1/ \nu_{\rm min}$ with $B_{\rm tot}$ is observed above $B_c$ for which the spin polarization is expected to be completed.
Similar behavior is found for $\rho_{xx}$ minima at $\nu \approx 1$ and 2 in the insulating regime as shown in Fig.~5.
\begin{figure}
\includegraphics{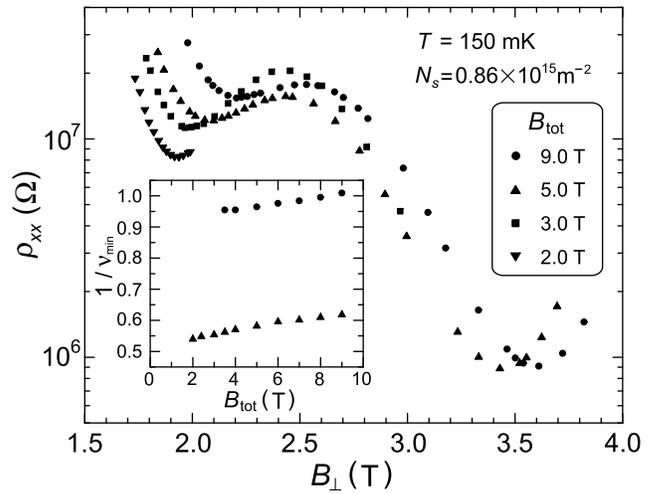}
\caption{
$B_\perp$-dependence of $\rho_{xx}$ at $T$ = 150 mK and $N_s  = 0.86 \times 10^{15} {\rm{m}}^{{\rm{ - 2}}} $ for various $B_{{\rm{tot}}} $.
$\rho_{xx}$ minima at $\nu \approx 1$ and 2 are observed.
The inset shows the position of the $\rho_{xx}$ minima. 
}
\end{figure}
We estimate that the change in $N_s$ due to the $B$-dependent shift of the chemical potential of the 2DES for a fixed gate voltage is negligible (in the order of $10^{-3} N_s$).
It seems possible that spin-down electrons occupy deep levels due to impurity potentials even at $B_{\rm tot} = B_c$ and are gradually released into non-trapped states with an up-spin at higher magnetic fields.
In the Si/SiGe sample, \cite{Okamoto2004} the increase in $1/ \nu_{\rm min}$ with $B_{\rm tot}$ is not observed above $B_c$.

The critical magnetic field $B_c$ is also obtained from a kink in a magnetoresistance curve in the in-plane magnetic field $B_\parallel$ since the $B_\parallel$-dependence of $\rho_{xx}$ is associated with the spin polarization. \cite{Okamoto1999,Okamoto2004,Okamoto2000,Tutuc2001,Tutuc2002}
Figure~6(a) shows the data in the insulating regime.
\begin{figure}[!t]
\includegraphics{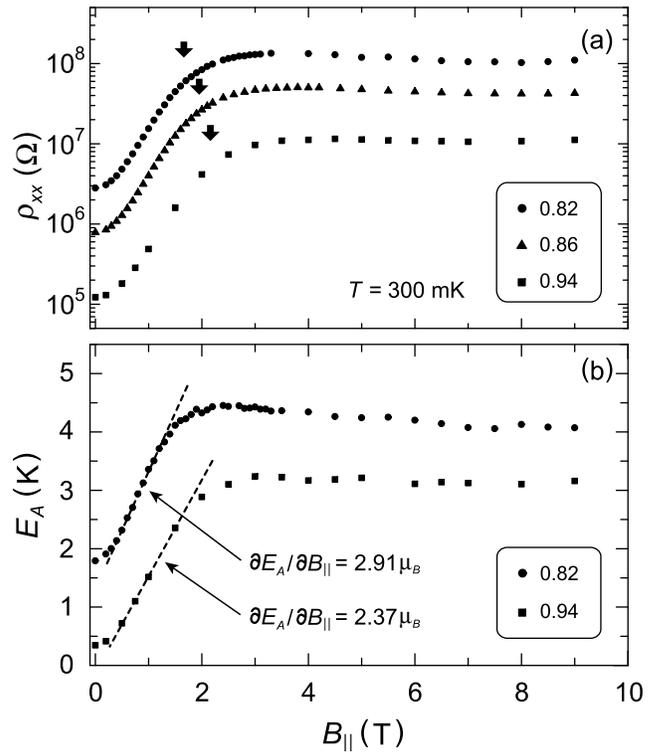}
\caption{
(a) In-plane magnetic field dependence of $\rho_{xx}$ for different $N_s $ at $T$ = 300 mK. The arrows indicate $B_c$ determined from $\rho_{xx}$ vs $B_\parallel$ data obtained for various temperatures.
(b) Activation energy determined from $T$-dependence of $\rho_{xx}$ for $N_s  = 0.82$ and $0.94 \times 10^{15}~{\rm{m}}^{-2}$. 
}
\end{figure}
While $\rho_{xx}$ has strong temperature dependence, the $B_\parallel$-dependence at a constant temperature shows a steep increase in the low $B_\parallel$ region and a saturation in the high $B_\parallel$ region.
The critical magnetic field determined from the magnetoresistance curve is consistent with that obtained from Fig.~4 as shown later in Fig.~7.
It also suggests that a ferromagnetic instability does not occur in the insulating regime. 
In Refs.~\onlinecite{Shashkin2001} and \onlinecite{Vitkalov2001}, the authors claimed that $B_c$ tends to vanish at an electron concentration close to the MIT at $B=0$ based on the analysis of the $B_\parallel$-dependence of $\rho_{xx}$ obtained in the metallic side.
However, the magnetoresistance in the far insulating regime was not studied. 

In Fig. 6(b), we show the activation energy $E_{A}$ determined from the Arrhenius temperature dependence of $\rho_{xx}$.
If we assume $E_{A}$ as an energy gap for an elementary excitation, the magnetization change due to the excitation can be obtained thermodynamically via the relation $\delta M =  - \partial E_{A} /\partial B_\parallel $.
A similar method was used for the study of the energy gap for the odd-integer quantized Hall states. \cite{Schmeller}
In our Si-MOSFET, the average distance of electrons from the Si/SiO${}_2$ interface \cite{Ando1982} is by one order of magnitude smaller ($\approx 3.8~{\rm nm}$) than the magnetic length $l_0 = (\hbar /eB_\parallel)^{1/2}$ in the low $B_\parallel$ region where the steep increase in $E_A$ is observed.
Since $B_\parallel$ does not couple to the orbital motion of electrons in this case, $\delta M$ should be attributed to electron spin flips.
The slope $\partial E_{A} /\partial B_\parallel $ in the low $B_\parallel$ region is somewhat larger than $+2\mu_{B}$ expected from a single spin flip.
Here, $\mu _B $ is the Bohr magneton and the bare $g$-factor in silicon is close to two.
The observed large $| \delta M |$ cannot be explained by a single particle picture.

Figure~7 shows $B_c$ as a function of $N_s$.
They are normalized by the non-interacting value $B_0 = \pi \hbar^2 N_s / 2 \mu_{B} m_b$ with $m_b = 0.19 m_0$.
\begin{figure}
\includegraphics{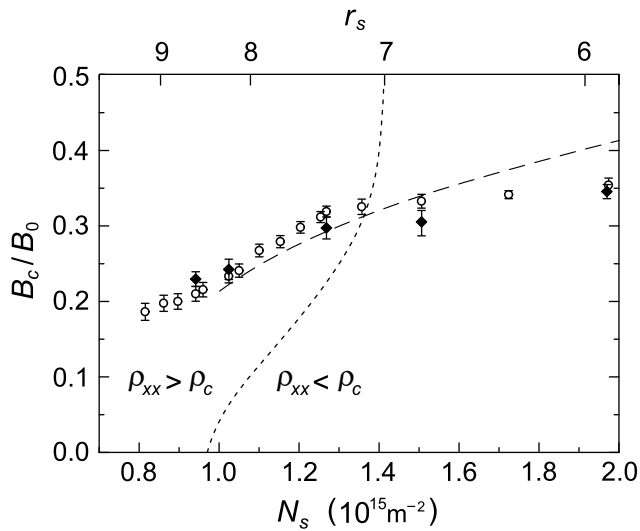}
\caption{
$B_c$ obtained from the $B_{\rm tot}$-dependence of $1/\nu_{\rm min}$ (closed diamonds) and that from the $B_\parallel$-dependence of $\rho_{xx}$ (open circles) are shown as a ratio to the non-interacting value $B_0$.
The dashed curve is an estimation from the spin susceptibility obtained by Pudalov {\it et al}. \cite{Pudalov2002} in the limit of small magnetic fields.
The dotted curve represents the values of $B_\parallel/B_0$ for a tentative boundary in the $B_\parallel$-$N_s$ plane determined from $\rho_{xx}$ at $T=150~{\rm mK}$. $\rho_c= 60~{\rm k}\Omega$ is the critical resistivity at $N_s =N_c$ and $B =0$.
}
\end{figure}
If $P$ increases linearly with $B_{\rm tot}$ below $B_c$, $B_c/B_0$ is equal to the inverse of the ratio of $\chi^*$ to non-interacting susceptibility $\chi_0$.
The dashed curve represents $\chi_0/\chi^*$ obtained from the data by Pudalov {\it et al.} \cite{Pudalov2002}
The dotted curve represents the values of $B_\parallel/B_0$ for a tentative boundary in the $B_\parallel$-$N_s$ plane on which $\rho_{xx}$ at $T=150~{\rm mK}$ is equal to the critical resistivity $\rho_c = 60~{\rm k}\Omega$ for the MIT at $B=0$. \cite{BMIT}
While the magnetic-field-induced MIT occurs in the intermediate $N_s$ range, $B_c/B_0$ is in good agreement with $\chi_0/\chi^*$ obtained in the metallic regime in the limit of small magnetic fields. \cite{Pudalov2002}
$B_c/B_0$ gradually decreases with decreasing $N_s$ (increasing $r_s$) in the whole $N_s$ range and the $N_s$-dependence shows no distinct anomaly.

\vspace{5mm}

\section{summary}

In summary, we have studied the low-temperature magnetoresistance in a high-mobility Si-MOSFET sample.
Even in the far insulating regime with $\rho_{xx} \sim 10^8~\Omega$, we observed a dip in $\rho_{xx}$ at $\nu \approx 4$ in the $B_\perp$-dependence at $B_{\rm tot} = 9~{\rm T}$.
The critical magnetic field for the onset of the full spin polarization determined from the $B_{\rm tot}$-dependence of the value of $B_\perp$ at the dip, which agrees with that from the magnetoresistance curve in the in-plane magnetic field, indicates that a ferromagnetic instability does not occur in the insulating regime.

It was found that $B_\perp$-dependence and $B_\parallel$-dependence of $\rho_{xx}$ at low temperature in the far insulating regime are the results of those of the activation energy $E_{A}$ in the Arrhenius temperature dependence.
However, the origins of the dip in $E_{A}$ at $\nu \approx {\rm integer}$ and the steep increase in $E_{A}$ with $B_\parallel$ for low $B_\parallel$ are not understood, while they might be associated with strong electron correlations.
Further investigations are required.

\section{acknowledgments}

We thank Dr. A. Yagi for providing us with the Si-MOSFET sample. 
This work is supported in part by Grants-in-Aid for Scientific Research from the Ministry of Education, Science, Sports and Culture, Japan.


\begin{thebibliography}{30}
\bibitem{BMIT}
Metallic behavior is not observed at high magnetic fields even in the low resistivity region. It is replaced by weak negative $T$-dependence.
\end{thebibliography}
\end{document}